\documentclass[12pt,a4paper,
 	       final]{scrartcl}

\usepackage{graphicx}%
\usepackage{ae,aecompl} 
\usepackage[english]{babel}
\usepackage[applemac]{inputenc} 
\usepackage[T1]{fontenc}
\usepackage[square,numbers,sort&compress]{natbib}

\usepackage{times} 

\usepackage[intlimits]{amsmath}
\usepackage{amsfonts}
\usepackage{amsthm}
\usepackage{mathrsfs}
\usepackage{stmaryrd}

\begin{document} 
 
\title{On the importance of hydrodynamic interactions in polyelectrolyte
electrophoresis}
\author{K Grass\thanks{Frankfurt Institute for Advanced Studies,
Goethe University, Ruth-Moufang-Strasse 1, D-60438 Frankfurt am Main, Germany, \texttt{grass@fias.uni-frankfurt.de}}
 \and C Holm$^{*,}$\thanks{Max-Planck Institute for Polymer Research, Ackermannweg 10,
D-55128 Mainz, Germany}}

\maketitle

\section*{Abstract}
The effect of hydrodynamic interactions on the free-solution electrophoresis of
polyelectrolytes is investigated with coarse-grained molecular dynamics
simulations. By comparing the results to simulations with
switched-off hydrodynamic interactions, we demonstrate their importance in
modelling the experimentally observed behaviour. In order to quantify the
hydrodynamic interactions between the polyelectrolyte and the
solution, we present a novel way to estimate its effective charge. We obtain
an effective friction that is different from the hydrodynamic
friction obtained from diffusion measurements. This effective friction is used
to explain the constant electrophoretic mobility for longer chains. To further
emphasize the importance of hydrodynamic interactions, we apply the model to
end-labeled free-solution electrophoresis.

PACS numbers: 82.35.Lr, 47.57.Ng, 82.35.Rs, 82.56.Lz, 82.56.Jn, 87.15.ap

Submitted to: J. Phys.: Condens. Matter

\section{Introduction}

Nowadays, electrophoretic separation methods are widely applied to study
polyelectrolytes (PEs) such as proteins, DNA and synthetic polymers
\cite{righetti96a,cottet05a,dolnik06a}. While there exist several theories
\cite{barrat96a,muthukumar96a,volkel95b,mohanty99a} that have  been
successfully used to describe qualitatively the experimentally observed
behaviour of various PEs, there are still many open problems to address.

Recent experiments on strongly charged flexible PEs, such as polystyrene
sulfonate (PSS) and single-stranded DNA (ssDNA) of well defined length have shown a
characteristic behaviour for the short chain free-solution mobility $\mu$
\cite{stellwagen02a,hoagland99a,cottet00b,stellwagen03a}: after an initial
increase of the mobility with increasing length, $\mu$ passes through a
maximum, and then decreases towards a constant mobility for long chains. 

The increase for short chains and the long-chain limit can be explained within
the theoretical approaches, but the origins of the maximum for intermediate
chains remain indistinct. To some extent this can be attributed to the
simplifying assumptions made in those models regarding the
interplay of the various interactions: the Coulomb interaction between the 
charged PE monomers and its counterions, the external electric field likewise
acting  on the charges, and the hydrodynamic interactions with the solvent.

To provide a fundamental understanding of the involved dynamics, we believe it
is mandatory to study the effects of these forces on a microscopic level,
thereby taking into account full electrostatic as well as hydrodynamic
interactions. Continuing the work of \cite{grass08a}, we use coarse-grained
molecular dynamics simulations to determine the transport coefficients and
structural properties of strong polyelectrolytes in free-solution
electrophoresis. We characterize the hydrodynamic interactions between solvent
and solute on a microscopic level, and determine the relevant length scale for
these interactions.

The article is structured as follows: in section~\ref{sec:model} we introduce
the simulation model and specify all relevant parameters. After comparing the
simulation results to experimental measurements of the PSS transport
coefficients for varying length (Section \ref{sec:pss}), we switch off
hydrodynamic interactions to characterize the behaviour in absence of them. In
section~\ref{sec:efffriction}, we introduce the concept for an effective
friction of the polyelectrolyte counterion compound with the surrounding
solvent. To quantify the effective friction, we propose a novel way to estimate
the effective charge. Finally, we show that the effective friction, measured in
this way, deviates from the hydrodynamic friction that can be obtained from
diffusion measurements, and that this deviation is the reason for the
observable length independent mobility of long polyelectrolyte chains. Finally,
in section~\ref{sec:elfse}, we demonstrate, how increasing the effective
friction by a drag label can restore size dependent mobility. We finish this
paper with concluding remarks.

\section{Simulation model}\label{sec:model}

We use a charged bead-spring model within the Espresso package
\cite{limbach06a}, to study the behaviour of flexible linear PEs of different
lengths. All parameters are given in reduced units with energy scale $kT =
1.0$ and relevant length scale $\sigma_0$ that is used to match the model to
a specific polyelectrolyte. For this paper, we chose $\sigma_0 = 2.5$ \AA, which
is the distance between two charges along a fully sulfonated PSS backbone. The
simulation time step is $\tau_0 = 0.01$.

The PE is comprised of $N$ negatively charged monomers carrying a
charge of $-1 e_0$, where $e_0$ is the elementary charge. 

The monomers are
connected by finitely extensible nonlinear elastic (FENE) bonds
$U_\text{FENE} = \frac{1}{2} k R^2 \ln \left( 1 - \left( \frac{r}{R} \right)^2
\right)$, 
with stiffness $k = 30$, and maximum extension $R = 1.5$, and with $r$ being the
distance between the interacting monomers \cite{soddeman01a}.
Additionally, a truncated Lennard-Jones or Weeks-Chandler-Anderson (WCA) potential 
$U_\text{LJ} = 4 \epsilon \left( \left( \frac{\sigma}{r}\right)^12 - \left(
\frac{\sigma}{r}\right)^6 \right)$,
with depth $\epsilon = 0.25$ and width $\sigma = 1$, is
used for excluded volume interactions \cite{weeks71a}.

For charge neutrality, $N$ monovalent counterions of charge $+1 e_0$ are added
that are subject to the same LJ potential. All particles have the same size.

The simulations are carried out under periodic boundary conditions in a
rectangular simulation box. The size $L$ of the box is varied to achieve a
constant monomer concentration $c_\mathrm{PE}$ and counterion
concentration $c_\mathrm{CI}$ independent of chain length. To match
experimental conditions with a PSS monomer concentration of 1 g/l or 5 mM, we
vary $L$ between 34 ($N=2$) and 89 ($N=32$). 

Full electrostatic interactions are calculated with the
particle-particle-particle mesh (P3M) algorithm \cite{deserno98a}. The Bjerrum
length $l_B = e_0^2 / 4 \pi \epsilon k T = 2.84$ in simulation units
corresponds to 7.1 \AA{} (the value for water at room temperature).

To account for hydrodynamic interactions (HI), we frictionally couple all
particles to a lattice Boltzmann fluid as detailed in \cite{ahlrichs99a}. The
modeled fluid has a kinematic viscosity $\nu = 3.0$, a fluid density $\rho =
0.864$, and is discretized by a grid with spacing $a = 1.0$. The coupling
parameter is $\Gamma_\mathrm{bare} = 20.0$.

In order to further characterize the impact of HI on the system's dynamics, we
compare to simulations with a simple Langevin thermostat that does not recover
long-range hydrodynamic interactions between the monomers, but only offers
local interactions with the solvent. We set the friction parameter $\Gamma_0
= 15.34$ to match the single particle mobility of the Langevin system to the
one with full HI.

\subsection{Determining transport coefficients} 

Using this model, we determine the single chain diffusion coefficient $D$ and
the electrophoretic mobility $\mu$ of the model PE from simulations without an
applied external field using the following Green-Kubo relations:

\begin{equation}\label{eq:diff}
    D = \frac{1}{3} \int_0^{\infty}  \langle\vec{v_c}(\tau)
	\cdot \vec{v_c}(0) \rangle d\tau
\end{equation}

\begin{equation}\label{eq:mob}
    \mu = \frac{1}{3 k_B T} \sum_i q_i \int_0^{\infty}
	\langle\vec{v_c}(\tau) \cdot \vec{v_i}(0) \rangle d\tau
\end{equation}

Here, $\vec{v_c}$ is the center of mass velocity of the PE, $\vec{v_i}$ is the
velocity of every charged particle in the system, i.e. monomers of the PE and
associated counterions, and $q_i$ is the corresponding charge. The ensemble
average ($\langle \ldots \rangle$) is taken over $10^4$ statistically
independent samples. The integrations use analytic fits for the slowly
converging long-time tails.

The advantage of using a Green-Kubo relation (2) for the electrophoretic mobility is that we can obtain both transport coefficients from the same simulation
trajectories at zero field. Doing so we avoid conformational changes to the
chain structures or the counterion distributions by an artificially high external electric field,
which is sometimes used in other simulations to separate the directed
electrophoretic motion from the Brownian motion within reasonable computing
time. This approach was successfully applied in simulations to determine the
electrophoretic mobility of charged colloids \cite{lobaskin07a,duenweg08a}.

\section{Results}\label{sec:results}

\subsection{Transport coefficients of PSS}\label{sec:pss}

\begin{figure}[htp] 
\begin{center}   
  \includegraphics[width=\columnwidth]{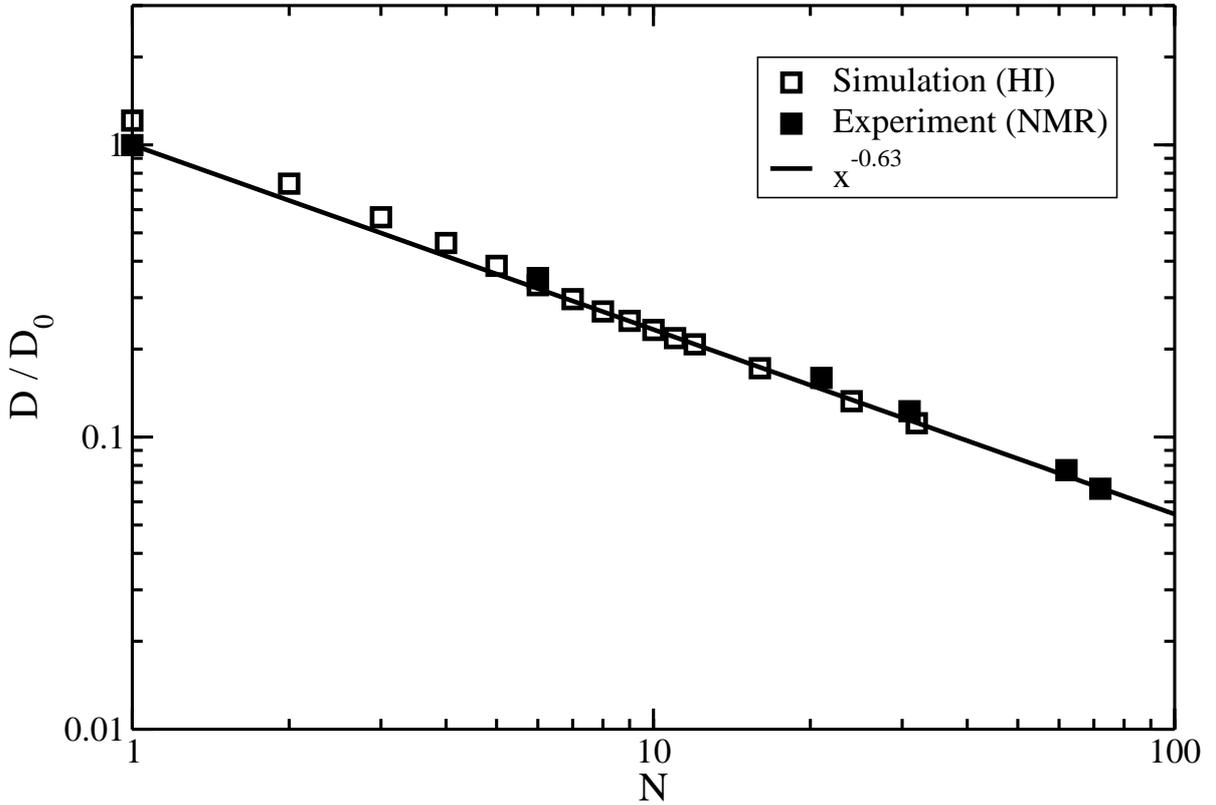}
  \caption{The normalized diffusion coefficient $D/D_0$ for PSS of different 
 lengths N as obtained by electrophoresis NMR agrees with the simulation
 results with full hydrodynamic interactions (HI).}
  \label{fig:diff}
  \end{center} 
\end{figure}  

\begin{figure}[htp]
\begin{center}
  \includegraphics[width=\columnwidth]{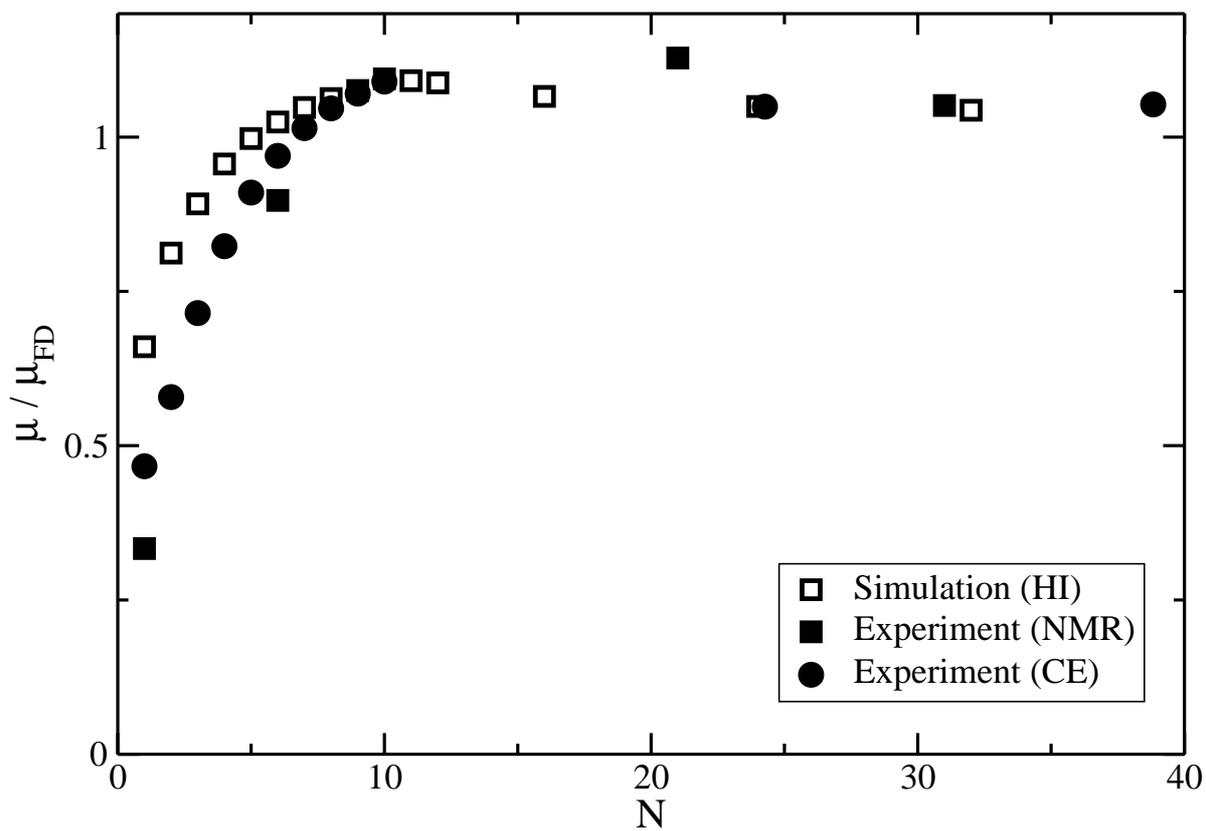}
  \caption[]{Including hydrodynamic interactions
  (HI), the normalized electrophoretic mobility $\mu / \mu_\mathrm{FD}$ as a
  function of the number of repeat units $N$ obtained in simulations reproduce
  a maximum for intermediate chains as well as the long-chain behaviour
  observed in experiments.}
  \label{fig:mob}
\end{center}
\end{figure}

In an earlier publication this year \cite{grass08a}, we compared the simulation
results to experimental data for short PSS obtained by two different
experimental methods, namely, capillary electrophoresis
\cite{nkodo01a,stellwagen02a} and pulsed field gradient or electrophoretic NMR
\cite{stejskal65a,scheler02a,gottwald03a,stilbs06a}\footnote{Please refer to
\cite{grass08a} for details on the experimental conditions.}. 

In Figure \ref{fig:diff}, we compare the diffusion coefficient $D$ determined
by simulations to the experimental results. Both results are in good
agreement and exhibit a power law scaling $D = D_0 N ^m$, with an scaling
exponent $m = 0.63 \pm 0.01$, which agrees with previous results
\cite{stellwagen03a,boehme03a,boehme07a,boehme07b}. Here, 
the simulated data is normalized by
$D_0 = 0.052$, the monomer diffusion as obtained by a power law fit, and the
experimental data by the monomer diffusion coefficient of $D_0 = 5.7 \cdot
10^{-10}$ m$^2/$s.

Figure \ref{fig:mob} shows the electrophoretic mobility $\mu$ normalized by
$\mu_\mathrm{FD}$, the constant (not length-dependend) value for long chains.
We compare the simulation results to two diffent experimental data sets. The
simulation results reproduce a maximum for intermediate chains as well as the
long-chain behaviour (the constant mobility for long chains) observed in
experiments, if the hydrodynamic interactions are properly accounted for.

\subsection{Absence of hydrodynamic interactions}\label{sec:nohd}

Without hydrodynamic interactions, the simulation fails to
describe the short chain behaviour and can not be mapped to the experimental data.

\begin{figure}[htp]
\begin{center}
  \includegraphics[width=\columnwidth]{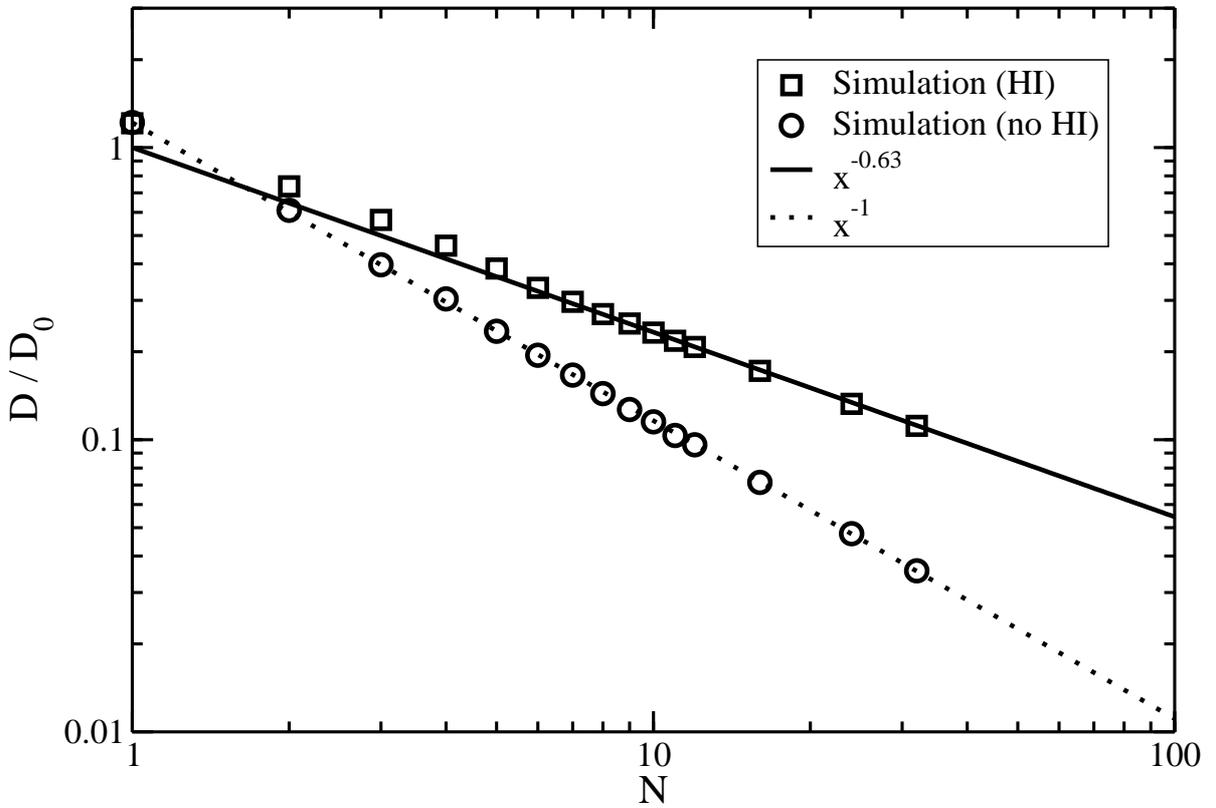}
  \caption[]{The normalized diffusion coefficient $D/D_0$ with and w
 ithout hydrodynamic interactions (HI). With HI the experimentally observed sca
ling is recovered; without HI the chain diffusion is Rousse-like.  }
  \label{fig:diffnohd}
\end{center}
\end{figure}

\begin{figure}[htp]
\begin{center}
  \includegraphics[width=\columnwidth]{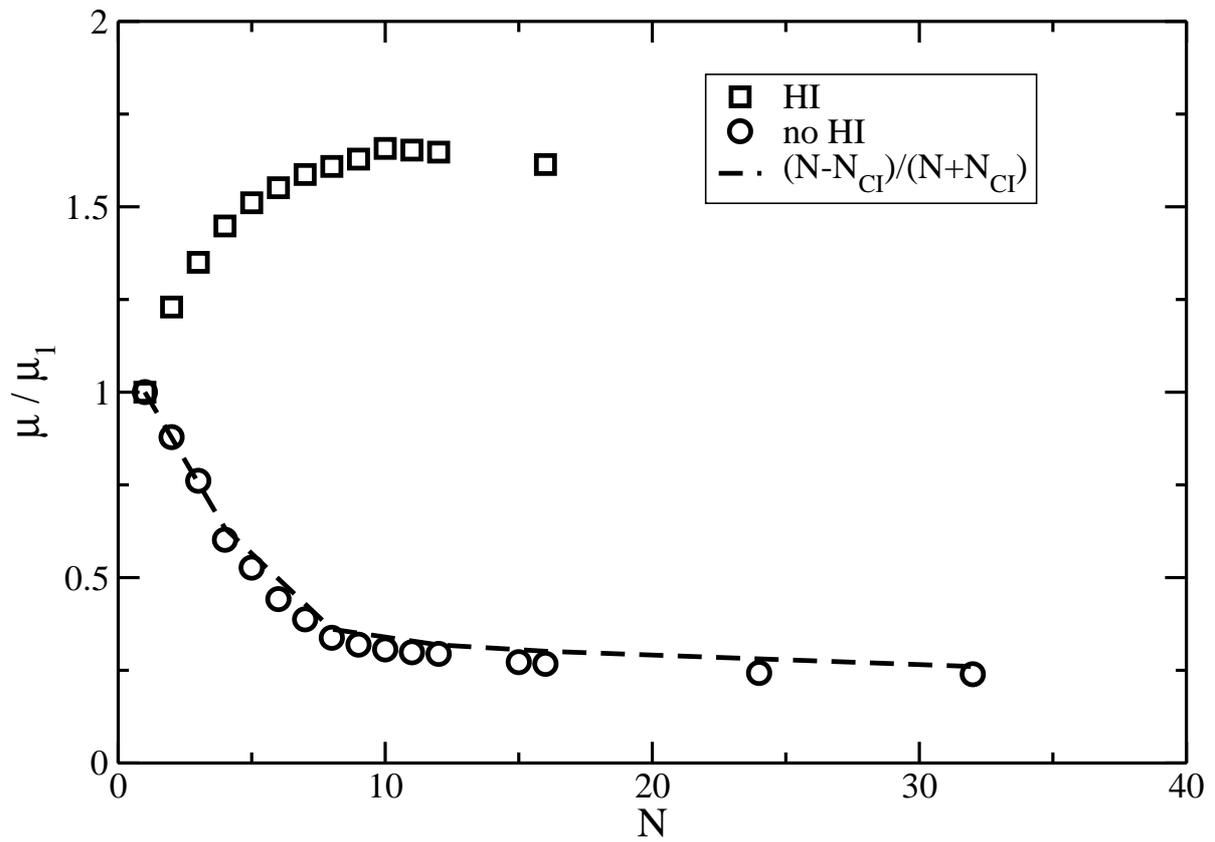}
  \caption[]{The normalized electrophoretic mobility without hydrodynamic
  interactions (HI) strongly deviates from the one obtained with HI. The
  plotted curve, equation (\ref{eq:langmob}), describes this observation by
  using the local force picture. }
  \label{fig:mobnohd}
\end{center}
\end{figure}

Figure \ref{fig:diffnohd} illustrates the difference in the scaling of the
diffusion coefficient with and without hydrodynamic interactions. For the
latter, Rousse behaviour is observed with a scaling exponent of $m = 1.02 \pm
0.02$ for the chain diffusion. Likewise, as shown in figure \ref{fig:mobnohd},
the neglect of hydrodynamic interactions leads to a decreasing
electrophoretic mobility for short chains and a constant value for long chains.
The same observations were made in a recent publication by
Frank and Winkler \cite{frank08a}. 

\subsection{Local force balance}\label{sec:localforcebalance}

The behaviour in absence of hydrodynamic interactions can be understood in a 
local force balance picture. A strongly-charged polyelectrolyte such as PSS is
surrounded by a cloud of oppositely charged counterions, some of which move
with the chain, thus forming a molecular compound with the PE that has a
reduced electric charge. This effect is also know as counterion condensation
\cite{manning98a}. Without hydrodynamic interactions, every particle of this
compound is subject to the same frictional force specified by the Langevin
thermostat. For a not to strong applied electrical field $E$, the
electrophoretic mobility can be obtained by $\mu = v/E$, where $v$ is the
steady-state velocity of the compound (PE and co-moving counterions). The two acting forces, the solvent
drag force $F_\mathrm{Solvent}$ and the electric driving force
$F_\mathrm{Field}$, are of equal magnitude and opposite direction. We now
define $N_\mathrm{CI}$ to be the number of co-moving counterions. This leads to
the following equation for the forces, where $\pm Q$ is the charge of the monomers
and the counterions, respectively.

\begin{equation}\label{eq:langsolvent}
  F_\mathrm{Solvent} = -\Gamma_0 v (N + N_\mathrm{CI})
\end{equation}
\begin{equation}\label{eq:langfield}
  F_\mathrm{Field} = Q E (N - N_\mathrm{CI})
\end{equation}

From this we can obtain the following expression for the electrophoretic
mobility in absence of hydrodynamic interactions, where $\mu_0 = 1/\Gamma_0$ is
the mobility of a single monomer:

\begin{equation}\label{eq:langmob}
	\frac{\mu}{\mu_0} = \frac{N-N_\mathrm{CI}}{N+N_\mathrm{CI}}
\end{equation}

For plotting (\ref{eq:langmob}) in figure \ref{fig:mobnohd}, we obtain
$N_\mathrm{CI}$ by counting the average number of counter-ions found within $2
\sigma_0$ of the chain. The local force picture successfully describes the
observed behaviour in absence of hydrodynamic interactions. 

In the linear response regime neither the chain structure nor the
counterion distribution is affected by the presence of hydrodynamic
interactions. Thus, the differences displayed in figures \ref{fig:diffnohd} and
\ref{fig:mobnohd} can only be attributed to different frictional forces
with the solvent acting on the PE-counterion compound. In the remainder of this
paper, we will investigate this in more detail.

\subsection{Introducing hydrodynamic friction}\label{sec:efffriction}

Figure \ref{fig:mobnohd} shows that the mobility strongly increases when
hydrodynamic interactions are taken into account. In other words, the
frictional forces the compound experiences from the solvent are reduced by the
hydrodynamic interactions between the monomers and the counterions. The solvent
interactions are no longer local, thus (\ref{eq:langsolvent}) and
(\ref{eq:langfield}) have to be modified:

\begin{equation}\label{eq:lbsolvent}
  F_\mathrm{Solvent} = -\Gamma_\mathrm{eff}(N,N_\mathrm{CI}) v
\end{equation}
\begin{equation}\label{eq:lbfield}
  F_\mathrm{Field} = Q_\mathrm{eff}(N,N_\mathrm{CI}) E
\end{equation}

Here, $\Gamma_\mathrm{eff}$ and $Q_\mathrm{eff}$ are the a priori unknown
effective friction and effective charge of the compound, that depend on the
degree of polymerization $N$ of the PE chain as well as on the number of
co-moving counterions $N_\mathrm{CI}$. Combining both equations leads to a
general expression for the electrophoretic mobility:

\begin{equation}\label{eq:lbmob}
  \mu = \frac{Q_\mathrm{eff}(N,N_\mathrm{CI})}{\Gamma_\mathrm{eff}(N,N_\mathrm{CI})}
\end{equation}

Since the general functional dependence of $\Gamma_\mathrm{eff}$ on $N$ and
$N_\mathrm{CI}$ in the presence of hydrodynamic interactions is non trivial, in
all but limiting cases, (\ref{eq:lbmob}) cannot be used to determine the
electrophoretic mobility analytically. However, it can be applied to determine
the effective friction from the electrophoretic mobility and the effective
charge: 
\begin{equation}\label{eq:gammaeff}
  \Gamma_\mathrm{eff} = Q_\mathrm{eff} / \mu.
\end{equation}

\subsection{Estimating the effective charge}

As before, the electrophoretic mobility is determined by (\ref{eq:mob}),
leaving the task to estimate the effective charge from the simulation. There
are several possible ways to determine the number of counterions that shield
the bare charge of the polyelectrolyte. In section~\ref{sec:localforcebalance},
we already used the straightforward way of counting the number of counterions
within a tube of radius $2 \sigma$ around the chain. 

\begin{equation}
	Q_\mathrm{eff}^{(1)} = N - N_\mathrm{CI}(d<2\sigma)
\end{equation}

This method has the draw back, that the threshold, i.e. the
size of the tube, is arbitrarily defined. Another frequently used way is the
inflection criterion \cite{belloni84a,deserno00a}, in which the integrated
radial counterion distribution is calculated and the inflection point used as a
threshold. 

Alternatively, we can rewrite (\ref{eq:langmob}) and use the electrophoretic
mobility without hydrodynamic interactions to calculate the effective charge.

\begin{equation}
	Q_\mathrm{eff}^{(2)} = N \left( 1- \frac{1-\mu \Gamma_0}{1+\mu \Gamma_0}\right)
\end{equation}

This definition of the effective charge does not have any free parameters
and since simulations without hydrodynamic interactions are computationally
inexpensive, we can determine $Q_\mathrm{eff}$ with high accuracy in this way.
Further analysis showed that $Q_\mathrm{eff}$ is insensitive to hydrodynamic
interactions in the linear response regime. 

\begin{figure}[htp]
\begin{center}
  \includegraphics[width=\columnwidth]{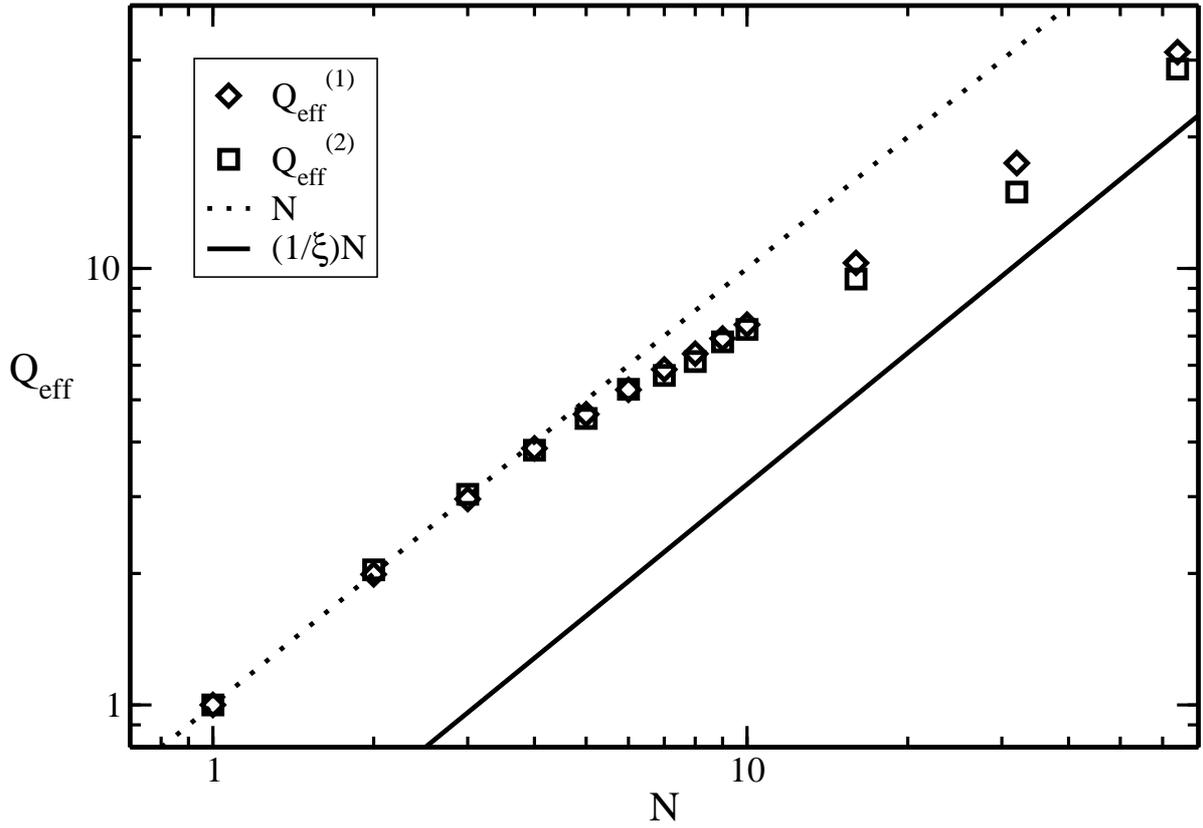}
  \caption[]{The effective charge $Q_\mathrm{eff}$ as a function of chain length
  for both estimators $Q_\mathrm{eff}^{(1)}$ and $Q_\mathrm{eff}^{(2)}$. The
  dotted line indicates the bare, unscreened charge of the polyelectrolyte, whereas
  the full line shows a prediction based on counterion condensation theory, with
  $\xi$ being the condensation parameter. }
  \label{fig:qeff}
\end{center}
\end{figure}

In figure \ref{fig:qeff}, we compare both estimators for the effective charge.
For short chains, both estimators agree and coincide with the bare, unscreened
charge of the polyelectrolyte. In this regime, no counterion condensation is
observed. 

For intermediate chains, the effective charge is reduced as it
deviates from the bare charge and tends towards the Manning prediction. Here,
the condensation parameter for the PSS system is $\xi = 3.12$. In this regime,
the simple estimator $Q_\mathrm{eff}^{(1)}$ measures a higher effective charge,
i.e.~not all condensed counterions that are included in the new estimator
$Q_\mathrm{eff}^{(2)}$ are taken into account.

The so defined effective charge estimators
$Q_\mathrm{eff}^{(1)}$ and $Q_\mathrm{eff}^{(2)}$ will now be used to quantify
the effective friction of the polyelectrolyte-counterion compound.

\subsection{Quantifying the effective friction}

\begin{figure}[htp]
\begin{center}
  \includegraphics[width=\columnwidth]{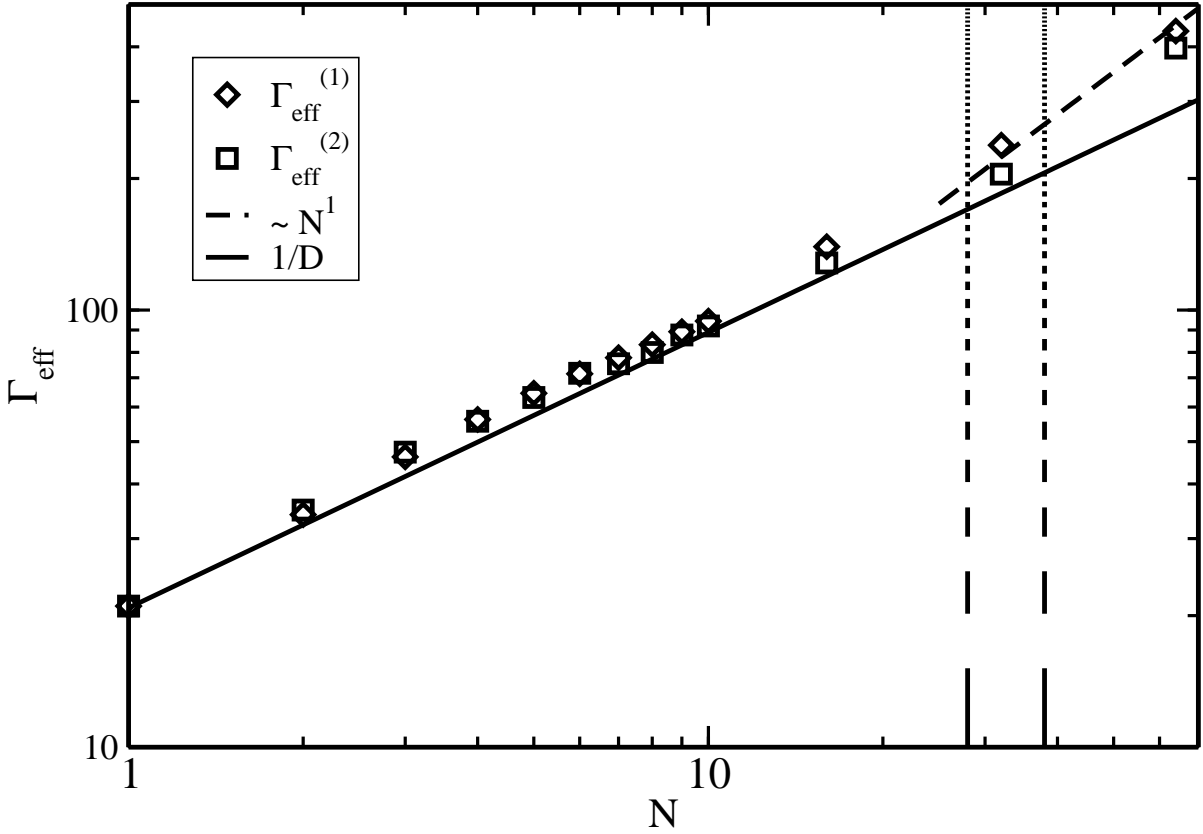}
  \caption[]{Effective friction of the polyelectrolyte and the co-moving
  counterions as abtained from (\ref{eq:gammaeff}) using $Q_\mathrm{eff}^{(1)}$
  and $Q_\mathrm{eff}^{(2)}$. For comparison, the inverse
  diffusion coefficient is displayed, which is a measure for the hydrodynamic
  friction if the Einstein equation (\ref{eq:einstein}) holds. The
  dashed lines indicate the approximate Debye screening length for the system.}
  \label{fig:gammaeff}
\end{center}
\end{figure}

\begin{figure}[htp]
\begin{center}
  \includegraphics[width=\columnwidth]{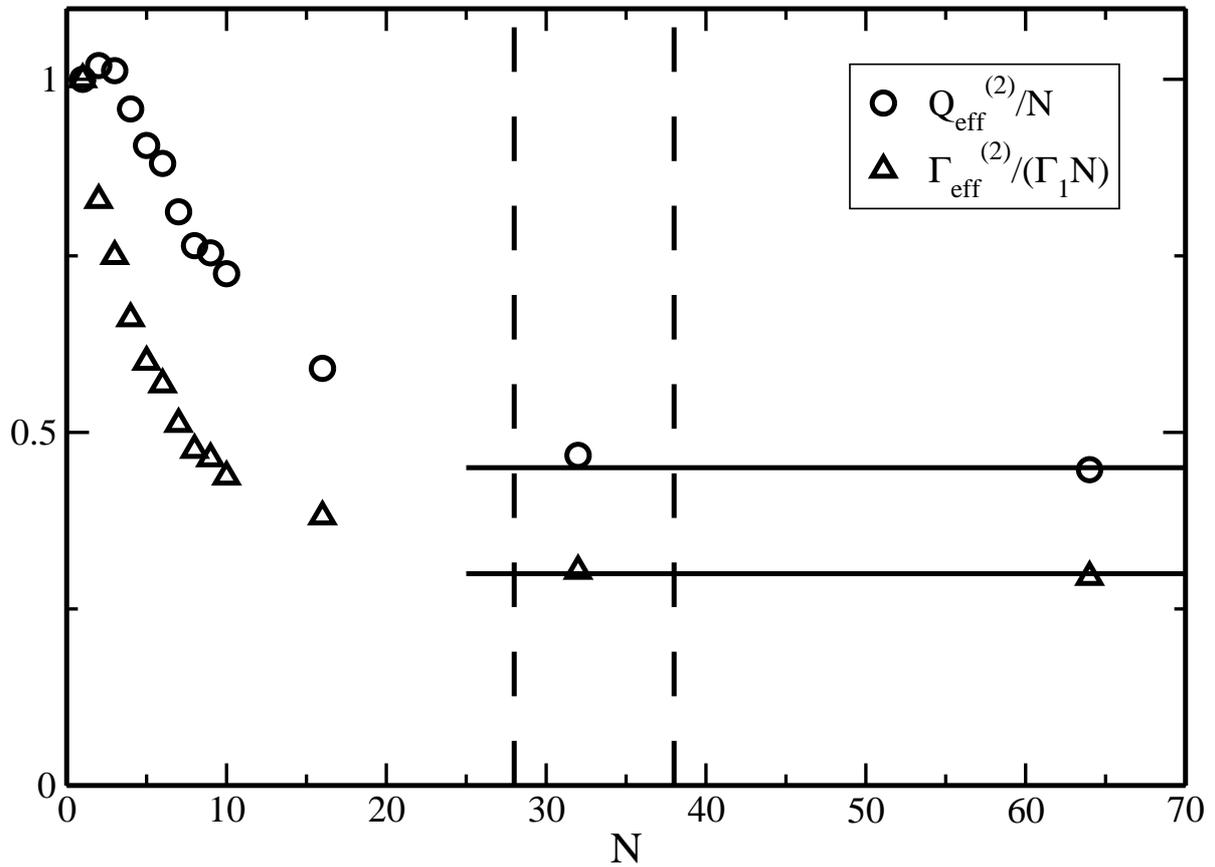}
  \caption{The effective charge and effective friction per monomer as a
  function of the chain length shows an initially stronger decrease of the
  friction compared to the charge which leads to the observed increase in
  mobility. For long chains, both quantities approach a constant value (the
  lines are visual guides only). Again the dashed lines indicate the range of
  electrostatic screening.}
  \label{fig:gammapermonomer}
\end{center}
\end{figure}

Figure \ref{fig:gammaeff} displays the effective friction of the compound in
dependence of the length of the PE chain as obtained from (\ref{eq:gammaeff})
using the estimators $Q_\mathrm{eff}^{(1)}$ and $Q_\mathrm{eff}^{(2)}$. We
compare the obtained result to the effective friction as it can be obtained via
Einstein equation from measurements of the diffusion of the PE:

\begin{equation}\label{eq:einstein}
	\Gamma_\mathrm{D} = \frac{1}{D}
\end{equation}

For chains with contour lengths of the order of the Debye length, a clear
deviation between $\Gamma_\mathrm{eff}$ and $\Gamma_\mathrm{D}$ is observed.
Beyond this length scale, the effective friction of the PE increases almost
linearly with chain length, where as the inverse diffusion scales with an
exponent of $m=0.63$ as shown in Section~\ref{sec:pss}.

In figure \ref{fig:gammapermonomer}, we compare the effective friction and the
effective charge per monomer. First, the effective friction shows a
stronger decrease than the effective charge. This leads to the observed
increase in mobility (see figure \ref{fig:mobnohd}). For longer chains, both
quantities approach a constant value. Again the relevant length scale
seems to be the Debye length, beyond which both quantities become constant
(i.e.~linear scaling of $Q_\mathrm{eff}$ and $\Gamma_\mathrm{eff}$).

This gives us a microscopic understanding of the effects, that lead to the
experimentally observed length independent mobility of long flexible
polyelectrolytes. The hydrodynamic interactions between polyelectrolyte and
counterions decrease the effective friction of the compound on length scales
smaller than the electrostatic screening length. For compounds of longer chains,
the hydrodynamic interactions are screened, and the friction
increases linear with the size of the PE. 

As an application of the presented model, we will now look at the a possible
way to increase the effective friction of the compound.

\subsection{Increasing the effective friction}\label{sec:elfse}

\begin{figure}[htp]
\begin{center}
  \includegraphics[width=\columnwidth]{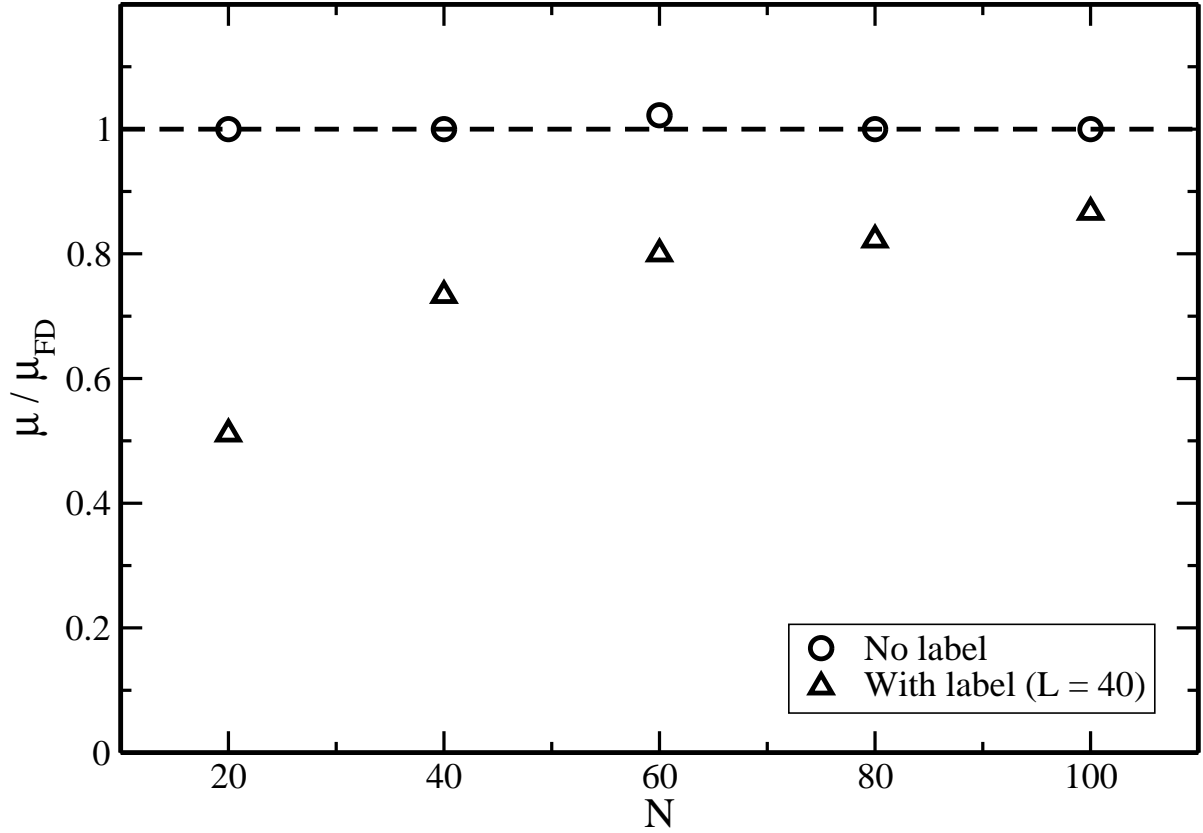}
  \caption[]{The mobility of a long flexible polyelectolyte without
  label is independent of the chain length $N$ (the line is a visual guide
  only). With a label of $L=40$ uncharged beads attached to the
  polyelectrolyte, the size dependence of the electrophoretic mobility can be
  recovered.}
  \label{fig:elfse}
\end{center}
\end{figure} 

As suggested by Mayer et.~al.~\cite{mayer94a,heller98a} attaching a drag label
("molecular parachute") to one end of the PE, increases the hydrodynamic friction by a constant amount independent of the chain length. This leads to a changed
scaling behaviour of the effective friction, which in turn can be seen in a
restoration of the size dependent mobility even for long PE chains.

In the simulations, the label consists of $L = 40$ uncharged monomers attached
to one end of the PE chain. The hydrodynamic interactions with the fluid cause an
additional force acting on the PE. Figure \ref{fig:elfse} demonstrates the
effectiveness of this method. 

In the future, we will use the toolkit described in this article to analyse
the ELFSE process in more detail.

\section{Conclusions}

We used a coarse-grained MD model to describe unlabeled and labeled
free-solution electrophoresis of flexible polyelectrolytes. The simulation
results are in good quantitative agreement with experimental data. We used
simulations without hydrodynamic interactions to efficiently measure the
effective charge of the polyelectrolyte as a function of the chain length. In a
further step, we quantified the effective friction of the compound formed by the
poylelectrolyte and co-moving counterions. Our results show that on a length
scale that is of the order of the Debye length for electrostatic screening the
friction becomes linear in terms of the chain length. This increase in friction
is exactly canceled out by the likewise linearly increasing effective charge,
leading to the well-known constant mobility for long flexible polyelectroyte
chains.

\section*{Acknowledgments}

We thank B.~D\"unweg, U.~Schiller, and G.~Slater for helpful remarks. Funds
from the Volkswagen Foundation, the DAAD, and DFG under Grant No.~TR6 are
gratefully acknowledged.

\end{document}